\documentstyle[prd,aps,floats,graphicx]{revtex}

\begin{document}
\preprint{ astro-ph/0012393}
\draft
%
\input epsf
\renewcommand{\topfraction}{0.99}
\twocolumn[\hsize\textwidth\columnwidth\hsize\csname 
@twocolumnfalse\endcsname

\title{A potential WIMP signature for the caustic ring halo model}
\author{Anne M.~Green}
\address{Astronomy Unit, School of Mathematical Sciences, Queen Mary 
University of London,\\ Mile End Road, London, E1 4NS,~~U.~~K.}
\date{\today} 
\maketitle
\begin{abstract}
Weakly Interacting Massive Particle (WIMP) direct detection event rate
calculations usually rely on fairly simple, essentially static,
analytic halo models. This is largely since the resolution of
numerical simulations is not yet large enough to allow the full
numerical calculation of the WIMP density and velocity distribution.
In this paper we study the direct detection rate, in particular its
energy dependence and annual modulation, for the caustic ring halo
model. In this model, which uses simple assumptions to model the
infall of dark matter onto the halo, the distribution of the cold dark
matter particles at the Earth's location has a series of peaks in
velocity space. We find that the recoil energy spectrum contains
distinctive steps and the sign of the annual modulation in the event
rate changes as a function of recoil energy. These effects provide a
potentially distinctive experimental signal.
\end{abstract}

\pacs{PACS numbers:  98.70.V, 98.80.C \hspace*{6.0cm}  astro-ph/0012393}

\vskip2pc]

\section{Introduction}

Weakly Interacting Massive Particle (WIMP) direct detection
experiments are just reaching the sensitivity required to probe the
interesting range of mass--cross-section parameter space where relic
neutralinos could constitute the dark matter.  The DAMA collaboration,
using a detector consisting of radiopure NaI crystal scintillators at
the Gran Sasso Laboratory, have reported the detection of a $4
\sigma$ annual modulation signal in their direct detection experiment,
consistent with WIMP scattering~\cite{dama,newdama}. Whilst this
result is somewhat controversial~\cite{concern,cdms} it illustrates
the potential of current and upcoming WIMP direct detection experiments.

Event rate calculations and detection strategies for particle physics
dark matter candidates are usually based on the assumption of a
standard Maxwellian halo model~\cite{am1,clas,jkg}. The standard
Maxwellian halo model has a number of deficiencies, in particular the
real halo contains substructure and is not perfectly spherical and
isotropic~\cite{Nbody}.  Since the resolution of numerical simulations
is not yet large enough to allow the full numerical calculation of the
WIMP density and velocity distribution, analytic, or at least
semi-analytic, models for the dark matter halo must be used. The
direct detection rate and in particular its annual modulation, which
occurs due to the Earth's motion, has been calculated for a range of
analytic non-standard halo
models~\cite{am1,ns1,br,vergados,ns2,uk,ecz,amg}. It has been found
that the region in the mass-cross section plane selected by the DAMA
data depends substantially on the halo model
assumed~\cite{br,ns2,amg}.

Analytic halo models usually assume an essentially static halo,
whereas in reality the halo is forming via the ongoing infall of
surrounding dark matter~\cite{gg}. The caustic ring halo model, which
arises from simple assumptions about the infall of dark matter onto
the halo, provides an analytic model of some features of the dark
matter distribution which may result from this accretion process. It
is therefore worthwhile to calculate the observational features which
the caustic ring halo model produces. The directional WIMP direct
detection rate, which would be probed by proposed experiments such as
DRIFT~\cite{drift}, has been calculated by Copi, Han and
Krauss~\cite{ck}, whilst Vergados~\cite{vergados2} has calculated the
total WIMP direct detection rate. In this paper we study the variation
of the differential direct detection rate, in particular its annual
modulation, with detector recoil energy.

\section{Caustic ring halo model}
\label{model}

Cold dark matter (CDM) particles are collisionless and have low
velocity dispersion ($<30 {\rm km s^{-1}}$) so that particles falling
onto an isolated galaxy are expected to oscillate in and out of the
galaxy a number of times before they are virialised by inhomogeneities
(such as molecular clouds, globular clusters and
stars)~\cite{is}. These non-virialised CDM flows lead to the formation
of caustic rings at the points where the particles with the most
angular momentum in a given inflow reach their point of closest
approach to the galactic centre, hence the name of the
model. Furthermore the distribution of the particles at any given
location is expected to have a series of peaks in velocity space,
corresponding to particles which are falling into the galaxy for the
first time and those which have fallen in and out a number of times
but have not yet been thermalized.  Whilst this model is obviously a
simplification of the hierarchical accretion process via which the
galactic halo forms, in particular the Milky Way is not an isolated
galaxy, the resolution of N-body simulations is not yet large enough to
resolve these sorts of features.

The velocities and densities at the Earth's location expected due to
these flows have been calculated using the self-similar infall
model~\cite{fgb} generalised to take into account the angular momentum
of the CDM particles~\cite{stw}. The Earth is located between the 4th
and 5th caustic rings and the velocity flows corresponding to these
two rings constitute roughly $ 30 \%$ of the local halo
density. Analysis of 32 extended galactic rotation curves has provided
some evidence for the 1st and 2nd caustic rings~\cite{ks}, whilst
analysis of an IRAS map of the galactic disk apparently reveals the
presence of the 5th ring~\cite{sikiras}.

The velocity distribution function of the velocity flows can be written as:
\begin{equation}
f({\bf v})= \Sigma_{j} \rho_{j} \delta ( {\bf v}-{\bf v}_j) \,,
\end{equation}
where $\rho_{j}$ and ${\bf v}_j$ are the density and velocity of the
$j$-th flow. Table I contains the most recently calculated values of
$\rho_{j}$ and ${\bf v}_j$~\cite{sikvalues} (note that there are two,
inward and outward, flows for each velocity peak). The total density
is $\rho_{0}=102 \, {\rm g  cm^{-3}}=0.57 \, {\rm GeV cm^{-3}}$, with the
velocity flows contributing $65\%$ of the total. We will assume that
the thermalized background distribution is a Maxwellian with velocity
dispersion $v_{0}=220 \, {\rm km s^{-1}}$.

\begin{table}
\begin{center}
\begin{tabular}{|c|c|c|c|c|}
j & $\rho_{j}$ ( $10^{-26} {\rm g cm^{-3}}$) & $v_{\phi}$ (${\rm km s^{-1}}$)
& ${v}_{{\rm z}}$  (${\rm km s^{-1}}$)  & $ {v}_{{\rm r}}$  (${\rm km s^{-1}}$) \\
 1   & 0.4 & 140& $\pm$ 605 & 0\\
 2    & 1.0  & 255 & $\pm$ 505 &  0\\
 3    & 2.0  & 350 & $\pm$ 390  & 0\\
 4    & 6.3 &  440 & $\pm$ 240 & 0\\
 5    & 9.2 &  440 & 0 & $\pm$ 190\\
 6 & 2.9 & 355 & 0 & $\pm$ 295 \\
 7 & 1.9 & 290 & 0& $\pm$ 330\\
 8 & 1.4 & 250 & 0 & $\pm$ 350\\
 9 & 1.1 & 215 & 0 & $\pm$ 355\\
 10 & 1.0 & 190 & 0 & $\pm$ 355\\
 11 & 0.9 & 170 & 0 & $\pm$ 355\\
 12  & 0.8 & 150 & 0 & $\pm$ 350\\
 13 &  0.7 & 135 & 0 & $\pm$ 345\\
 14 &  0.6 & 120 & 0  & $\pm$ 340\\ 
 15 & 0.6 & 110  & 0 & $\pm$ 330 \\
 16 & 0.55 & 100 & 0 & $\pm$ 325 \\
 17 & 0.50 & 90 & 0 & $\pm$ 320 \\
 18 & 0.50 & 85 & 0 & $\pm$ 310 \\
 19 & 0.45 & 80 & 0  & $\pm$ 305\\
 20 & 0.45 & 75 & 0  & $\pm$ 300\\
\end{tabular}
\end{center}
\caption[dfmf]{\label{d} The density and velocity components, in the
rest frame of the galaxy, of the velocity flows.}
\end{table}

\section{Annual modulation signal}
The WIMP detection rate depends on the speed distribution of the WIMPs
in the rest frame of the detector, $f_{v}$. This is found from the
halo velocity distribution, $f({\bf v})$ by making a Galilean
transformation ${{\bf v}} \rightarrow {\tilde{\bf v}}= {\bf v} - {\bf
v_{{\rm e}}}$, where ${\bf v_{{\rm e}}}$ is the Earth's velocity
relative to the galactic rest frame, and then integrating over the
angular distribution.  In galactic co-ordinates the axis of the
ecliptic lies very close to the $\phi-z$ plane and is inclined at an
angle $\gamma \approx 29.80^{\circ}$ to the $\phi-r$ plane. Including
all components of the Earth's motion, not just that parallel to the
galactic rotation~\cite{vergados}:
\begin{eqnarray}
{\bf v_{{\rm e}}}& =& v_{1} \sin \alpha \, {\hat r}  +  \nonumber \\
                 &&    (v_{0} +v_{1} \cos \alpha
                    \sin \gamma ) \,  {\hat \phi} -  v_{1} \cos \alpha
                   \cos \gamma  \, {\hat z}\,,   
\end{eqnarray}   
where $v_{0} \approx 232 {\rm km s^{-1}}$ is the speed of the sun with
respect to the galactic rest frame, $v_{1} \approx 30 {\rm km s^{-1}}
$ is the orbital speed of the Earth around the Sun and 
$\alpha= 2 \pi (t-t_{{\rm
0}})/ T$, with $T=1$ year and $t_{{\rm 0}} \sim 153$ days (June 2nd), 
when the component of the Earth's velocity parallel to the Sun's motion 
is largest. 

In the range of masses and interaction cross sections accessible to
current direct detection experiments the best motivated WIMP candidate
is the neutralino, for which the event rate is dominated by the scalar
contribution. The differential event rate simplifies to (see
e.g. Refs.~\cite{jkg,amg} for details):
\begin{equation}
\frac{{\rm d} R}{{\rm d}E} = \xi \sigma_{{\rm p}} 
              \left[ \frac{\rho_{0.3}}{\sqrt{\pi} v_{0}}
             \frac{ (m_{{\rm p}}+ m_{\chi})^2}{m_{{\rm p}}^2 m_{\chi}^3}
             A^2 T(E) F^2(E) \right] \,,
\end{equation}
where it is conventional to normalise the local WIMP density,
$\rho_{\chi}$, to a fiducial value $\rho_{0.3} =0.3 \, {\rm GeV
cm^{-3}}$, such that $\xi=\rho_{\chi} / \rho_{0.3}$, $E$ is the energy
deposited in the detector, $A$ is the atomic number of the detector
nuclei, $F(E)$ is the detector form factor (the Saxon Woods form
factor is used for $I$ whilst that of ${\rm Na}$ is taken to be unity,
see e.g. Ref.~\cite{br}) and $T(E)$ is defined as~\cite{jkg}
\begin{equation}
\label{tq}
T(E)=\frac{\sqrt{\pi} v_{0}}{2} \int^{\infty}_{v_{{\rm min}}} 
            \frac{f_{v}}{v} {\rm d}v \,,
\end{equation}
where   $v_{{\rm
min}}$ is the minimum detectable WIMP velocity
\begin{equation}
v_{{\rm min}}=\left( \frac{ E (m_{\chi}+m_{A})^2}{2 m_{\chi}^2 m_{A}} 
             \right)^{1/2} \,,
\end{equation}
$m_{\chi}$ is the WIMP mass and $m_{A}$ is the atomic mass of the
target nuclei.

In order to compare the theoretical signal with that observed we need
to take into account the response of the detector. The electron
equivalent energy, $E_{{\rm ee}}$, which is actually measured is a
fixed fraction of the recoil energy: $E_{{\rm ee}}= q_{{\rm A}}
E$. The quenching factors for I and Na are $ q_{{\rm I}}=0.09$ and
$q_{{\rm Na}}=0.30$ respectively~\cite{q}. The energy resolution of
the detector~\cite{ns1} is already taken into account in the data
released by the DAMA collaboration.

\begin{table}
\begin{center}
\begin{tabular}{|c|c|c|c|c|}
j & $\rho_{j}$ ( $10^{-26} {\rm g cm^{-3}}$)  & $v_{\phi}$ (${\rm km 
s^{-1}}$)  
& $\tilde{v}_{\phi}$ (${\rm km s^{-1}}$)   & $\tilde{v}_{{\rm tot}}$
(${\rm km s^{-1}}$) \\
 1   & 0.4 & 140&-104 (-78)& 609 (605)\\
 2    & 1.0  & 255 &11 (37)&  505 (506)\\
 3    & 2.0  & 350 &106 (132)& 409 (416)\\
 4    & 6.3 &  440 &196 (222)& 310 (327)\\
 5    & 9.2 &  440 &196 (222)& 273 (292)\\
 6 & 2.9 & 355 &111 (137)&311 (321)\\
 7 & 1.9 & 290 & 46 (72) & 333 (338)\\
 8 & 1.4 & 250 & 6 (32) & 350 (351)\\
 9 & 1.1 & 215 & -29 (-3)& 356 (355)\\
 10 & 1.0 & 190 & -54 (-28)& 359 (356)\\
 11 & 0.9 & 170 &-74 (-48)& 363 (358)\\
 12  & 0.8 & 150 &-94 (-68)& 362 (357)\\
 13 &  0.7 & 135 &-109 (-83)& 362 (355)\\
 14 &  0.6 & 120 &-124 (-98)& 362 (354)\\ 
 15 & 0.6 & 110  &-134  (-108)&356 (347) \\
 16 & 0.55 & 100 &-144  (-118)&355 (346) \\
 17 & 0.50 & 90 &-154 (-128) &355 (345) \\
 18 & 0.50 & 85 & -159 (-133)& 348 (337)\\
 19 & 0.45 & 80 & -164  (-138)&346 (335)\\
 20 & 0.45 & 75 & -169  (-143)& 344 (332)\\
\end{tabular}
\end{center}
\caption[tab2]{\label{tab2} The density, $\phi$ velocity component in
the rest frame of the galaxy $v_{\phi}$, $\phi$ velocity component in
the rest frame of the Earth $\tilde{v}_{\phi}$, and total speed in
the rest frame of the Earth $\tilde{v}_{{\rm tot}}$, of the caustic flows in June, when
$\alpha=0$ (and in December when $\alpha=\pi$).}
\end{table}

The expected experimental spectrum per energy bin for the DAMA
collaboration set-up is then given by~\cite{br}
\begin{eqnarray}
\frac{\Delta R}{\Delta E} (E) &=& r_{{\rm Na}} \int_{E/q_{{\rm
             na}}}^{(E+ \Delta E)/q_{{\rm na}}} \frac{{\rm d}R_{{\rm
             Na}}  }{{\rm d} E_{{\rm ee}}} (E_{{\rm ee}})  \frac{{\rm d}
             E_{{\rm ee}}}{ \Delta E} \nonumber \\ &&  + \, r_{{\rm I}}
              \int_{E/q_{{\rm
             I}}}^{(E+ \Delta E)/q_{{\rm I}}} \frac{{\rm d}R_{{\rm
             I}}  }{{\rm d} E_{{\rm ee}}} (E_{{\rm ee}})  \frac{{\rm d}
             E_{{\rm ee}}}{ \Delta E} \,,
\end{eqnarray}
where $r_{{\rm Na}}=0.153$ and $r_{{\rm I}}=0.847$ are the mass
fractions of Na and I respectively.
Since $v_{0} \gg v_{1}$ the differential event rate in the k-th energy bin
can be expanded in a Taylor series in $\cos \alpha$~\cite{clas}:
\begin{equation} 
\frac{\Delta R}{\Delta E_{{\rm }}}(E_{{\rm k}}) \approx S_{{\rm 0, k}} 
           + S_{{\rm m, k}} \cos \alpha \,.
\end{equation}

\section{Results}

Whilst all 3 components of the Earth's velocity need to be included to
calculate the annual modulation signal accurately, the signal is
largely determined by the component in the galactic plane~\cite{clas}: 
\begin{equation}
v_{{\rm e,} \phi}= v_{{\rm circ}} \left[ 1.05 + 0.06 \cos{\alpha} 
\right] \,,
\end{equation}
where $v_{{\rm circ}}=220 \, {\rm km s^{-1}}$ is the local circular
velocity about the galactic centre. Before presenting the results of
a numerical calculation, using all three components of the Earths'
velocity, we will carry out a simple analytic calculation, using only
the component in the galactic plane, in order to elucidate the
physical origin of the variation in $T(E)$.

\begin{figure}
\centering
\leavevmode\epsfysize=6.5cm \epsfbox{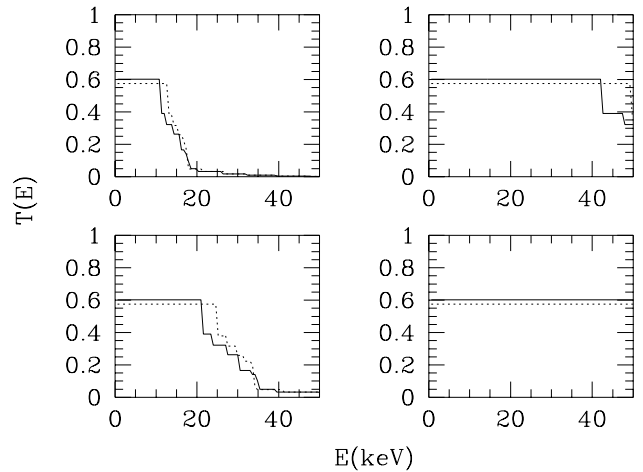}\\
\caption[fig1]{\label{fig1} The value of $T(E)$ in June (solid line)
and December (dotted line) due to the velocity flows alone for four
values of the WIMP mass $m_{\chi}=30,50,100, 200$ GeV, (top left,
bottom left, top right and bottom right respectively).}
\end{figure}

In June when $\alpha=0$
\begin{equation}
v_{{\rm e},\phi}=1.11 \times v_{\odot}=244.2 {\rm km s^{-1}} \,,
\end{equation}
whilst in December when $\alpha=\pi$
\begin{equation}
v_{{\rm e},\phi}=0.99 \times v_{\odot}=217.8 {\rm km s^{-1}} \,.
\end{equation}
Table 2 contains the density, $\phi$-velocity component, in the rest
frames of the galaxy and Earth, and the total velocity in the rest
frame of the Earth of the velocity flows for $\alpha=0$ and $\pi$. In
both cases the total density in flows with negative $v_{\phi}$
(incident from the forward direction) is 17.1 $\times 10^{-26} \, {\rm
g cm^{-3}}$ whilst the total density in flows with positive $v_{\phi}$
(incident from the backward direction) is 49.4$ \times 10^{-26} \,
{\rm g cm^{-3}}$ i.e. there are more WIMPs incident from backwards
than forwards as found by Copi, Han and Krauss~\cite{ck}. This is the
opposite of the directional signal produced by a pure Maxwellian halo.

\begin{figure}
\centering
\leavevmode\epsfysize=6.5cm \epsfbox{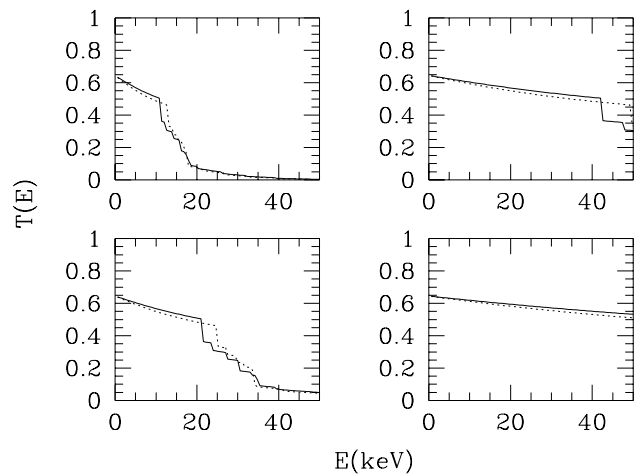}\\
\caption[fig2]{\label{fig2} The value of $T(E)$ in June (solid line) 
and December (dotted line) for a halo model with caustics plus an isothermal 
background, for four values of 
the WIMP mass $m_{\chi}=30,50,100, 200$ GeV, (as before).}
\end{figure}

In order to illustrate how the variations due to the caustics are
smoothed out by the isothermal background we plot $T(E)$, as a
function of $E$, for a ${\rm Ge}^{76}$ detector in Fig. 1 for the
velocity flows alone, and in Fig. 2 for the complete halo model
described above in Sec.~\ref{model}, where the velocity flows
contribute $65\%$ of the local density with the remaining $35\%$ in an
isothermal background. Values for other monatomic detectors can be
found by rescaling the x-axis by $m_{{\rm A}}/(m_{{\rm A}} +
m_{\chi})^{2}$.  

For a pure Maxwellian halo the signal is largest in
December for small recoil energies, switching to become largest in
June as the recoil energy is increased~\cite{uk}. The signal for the
velocity flows alone is more complicated. The contribution of the
$j$-th velocity flow to $T(E)$ is proportional to $\rho_{j}/
\tilde{v}_{{\rm tot}}$ if $ \tilde{v}_{{\rm tot}} > v_{{\rm min}}$ and
is zero otherwise.  The contribution of the high density flows to the
signal is largest in June, since their $\tilde{v}_{{\rm tot}}$ is
smaller in June than in December. Therefore at low energies, where all
the velocity flows contribute to the signal, the signal is largest in
June. A given high density velocity flow stops contributing to the
signal, $ \tilde{v}_{{\rm tot}} < v_{{\rm min}}$, for smaller $v_{{\rm
min}}$, or equivalently $E$, in June compared to December however.
This means that the steplike decreases in $T(E)$, which arise when a
given flow stops contributing to the signal, occur at lower energies
in June than in September (see Fig. 1). In other words for some range
of recoil energies a given flow contributes to the signal in December
but not in June.  At large recoil energies only the low density flows,
with high total velocity, can contribute and consequently the signal
is far smaller than at low recoil energies.  The lower density flows
have negative $\tilde{v}_{\phi}$ and, in contrast to the high density
flows, have larger speeds in June than in December, so that at high
recoil energies the contribution due to a given flow is slightly
larger in December. As $m_{\chi}$ is increased the variations in
$T(E)$ are moved to higher $E$. The presence of a Maxwellian background
smoothes the stepped variations in the signal produced by the flows,
but they are still discernible and if detected would provide a
distinctive indication of the presence of velocity flows.

In Fig. 3 we plot the differential event rate, ${\rm d}R/ {\rm d}E$,
for the velocity flows plus isothermal background model and also for a
pure Maxwellian halo, for a ${\rm NaI}$ detector using the best fit
values of the WIMP mass and cross-section found by the DAMA
collaboration, $m_{\chi}=54$ GeV $ \xi \sigma_{{\rm p}}=4 \times
10^{-6}$ pb. 

\begin{figure}
\centering
\leavevmode\epsfysize=6.5cm \epsfbox{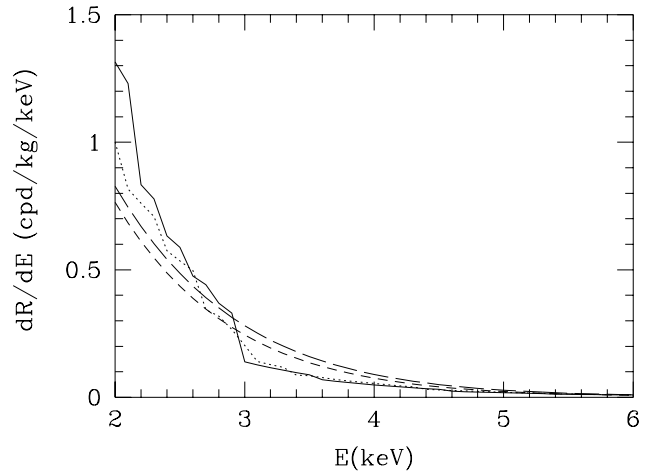}\\
\caption[fig3]{\label{fig3} The differential event rate ${\rm d}R/
{\rm d}E$ in June (solid line) and December (dotted line) for the
caustic ring halo model with velocity flows plus a Maxwellian
background as described in the text, and for a pure Maxwellian halo
(June-long dashed line, December-short dashed line), for WIMP mass
$m_{\chi}=54$ GeV and cross-section $ \xi \sigma_{{\rm p}}=4 \times
10^{-6}$ pb, as found by the DAMA collaboration, for a NaI detector.}
\end{figure}

Brhlik and Roszkowski~\cite{br} have devised a technique for comparing
the experimental data released by DAMA with theoretical predictions
for the annual modulation signal, in the absence of detailed
information about the experimental set-up, such as the efficiency of
each NaI crystal. Their technique, which is effectively a
least-squares comparison of the experimental data with the theoretical
predictions, has been used to examine the region of
mass--cross-section parameter space compatible with the DAMA results
for various simple non-standard, but close to Maxwellian, halo
models~\cite{br,amg}. The best fit values and errors for $S_{{\rm 0,
k}}$ and $S_{{\rm m, k}}$ released by DAMA are calculated under the
assumption that the recoil energy spectrum has the shape expected from
a Maxwellian velocity distribution. Therefore whilst Brhlik and
Roszkowski's technique can be used for halo models which produce
recoil energy spectra close to that produced by a Maxwellian, it can
not be applied to the caustic flow model. Furthermore Gelmini and
Gondolo have recently found that at low recoil energies the annual
modulation produced by the flows is poorly approximated by a
sinusoidal~\cite{gelmgond}.

We can therefore only make a qualitative discussion of the effect of
the presence of velocity flows on the DAMA allowed region.  Brhlik and
Roszkowski~\cite{br} found that, for a pure Maxwellian halo, the
cut-off at large WIMP masses in the allowed region is determined by
the time-dependent part of the signal (i.e. $S_{{\rm m,k}}^{{\rm
th}}$), whilst the lower limit on the WIMP mass depends on both the
time independent and dependent parts of the signal. When a velocity
flow component is added to the Maxwellian background the recoil energy
spectrum falls off less rapidly with increasing recoil energy for
large WIMP masses, whilst for smaller WIMP masses the recoil energy
spectrum falls off more rapidly with increasing energy (see Figs.~2
and 3). This suggests that the range of WIMP masses compatible with
the energy distribution observed by DAMA would be likely to be smaller
for the caustic ring model than for a pure Maxwellian halo.  The
allowed region obviously also depends on the magnitude and sign of the
annual modulation. For the caustic flow model the sign of the
modulation is opposite to that observed by DAMA for some, WIMP mass
dependent, ranges of recoil energy, however since the experimental
data is binned in 1keV bins this may not prevent the velocity flow
model being consistent with the DAMA data. It is possible though that
the distinctive effects of the velocity flows on the recoil energy
spectrum and on the sign of the annual modulation could lead to limits
on the allowed fraction of the local halo density in velocity flows.

For the purpose of estimating WIMP direct detection rates the
assumption of a standard Maxwellian halo is certainly reasonable.  Now
that experiments are reaching the region of parameter space populated
by supersymmetric models, and in the case of DAMA claiming a positive
signal, it is important to extend the theoretical analysis to more
sophisticated, and hopefully more realistic, halo models.  This process
will be facilitated by the public release of data in a form subject to
the minimum number of theoretical assumptions possible.

\section{Conclusions}
In this paper we have studied the WIMP direct detection signal, in
particular its annual modulation, for the caustic ring halo model. In
this model the WIMP distribution at the Earth's location has a series
of peaks in velocity space, corresponding to particles which are
falling into the galaxy for the first time and those which have fallen
in and out a number of times but have not yet been thermalized. These
peaks produce a distinctive imprint in the differential event rate,
with the sign of the annual modulation (i.e. whether the event rate is
larger in June or December) changing with detector recoil energy. The
presence of an isothermal background component to the halo smoothes
out the sharp changes in the differential event rate produced by the
velocity flows but the distinctive changes in the sign of the annual
modulation remain potentially discernible.

Finally we discussed the compatibility of this model with the results
of the DAMA experiment. The recoil energy spectrum varies more rapidly
with WIMP mass than that produced by a standard Maxwellian halo,
whilst for some recoil energies the annual modulation signal has the
opposite sign to that observed by DAMA. These effects suggest that for
this model the region of WIMP mass--cross-section parameter space
compatible with the DAMA data would be smaller than for the standard
Maxwellian halo model. In addition it may be possible, via a full
likelihood analysis, to constrain the fraction of the local halo
density in velocity flows. This illustrates that if a significant
component of the galactic dark matter is composed of WIMPs, then WIMP
direct detection experiments with fine grained directional and energy
resolution may be able to probe the local galactic structure,
complementing the information which indirect detection
experiments~\cite{idd} would be able provide on larger scales.

\section*{Acknowledgements}

A.M.G.~was supported by PPARC and acknowledges use of the Starlink
computer system at QMW. A.M.G~thanks Craig Copi, Simon Goodwin and
especially Pierre Sikivie for useful discussions, and the anonymous
referee for valuable comments.


\begin{references}
\bibitem{dama} R. Bernabei et. al. Phys. Lett. {\bf B389}, 757 (1996); 
               ibid {\bf B408}, 439 (1997); ibid {\bf B424}, 195 (1998); 
                 ibid {\bf B450}, 448 (1999).
\bibitem{newdama} R. Bernabei et. al. Phys. 
               Lett. {\bf B480}, 23 (2000).
\bibitem{concern} G. Gerbier, J. Mallet, L. Mosca and C. Tao, 
                astro-ph/9710181; astro-ph/9902194.
\bibitem{cdms} R. Abusaidi et. al. (CDMS Collaboration), Nucl. I
               nstrum. Meth. {\bf A444}
               345  (2000); Phys. Rev. Lett. {\bf 84}, 5699 (2000).
\bibitem{am1}  A. K. Drukier, K. Freese and D. N. Spergel, Phys. Rev. 
            D {\bf 33}, 3495 (1986). 
\bibitem{clas} K. Freese, J. Frieman and A. Gould, 
              Phys. Rev. D {\bf 37}, 3388 (1988).
\bibitem{jkg} G. Jungman, M. Kamionkowski and K. Griest, Phys. Rep. 267, 
               195 (1996).
\bibitem{Nbody} J. F. Navarro, C. S. Frenk and S. D. M. White, Astophys.
             J. {\bf 462}, 563 (1996); B. Moore et. al. Mon. Not. R. 
            Astron. Soc. {\bf 310} (1999), 1147; A. V. Kravtsov et. al. 
            Astrophys. J. {\bf 502}, 48 (1998).
\bibitem{ns1} F. Donato, N. Fornengo and S. Scopel, Astropart. Phys. {\bf 9},
              247 (1998). 
\bibitem{br} M. Brhlik and L. Roskzkowski, Phys. Lett. 
              {\bf B464}, 303 (1999).
\bibitem{vergados} J. D. Vergados, Phys. Rev. Lett. 
               {\bf 83}, 3597 (1999); Phys. Rev. D {\bf 62}, 023519 (2000).
\bibitem{ns2} P. Belli et. al. Phys. Rev. D {\bf 61}, 023512 (2000).
\bibitem{uk} P. Ullio and M. Kamionkowski, hep-ph/0006183.
\bibitem{ecz}  N. W. Evans, C. M. Carollo and P. T. de Zeeuw, 
           Mon. Not. R. Astron. Soc. {\bf 318}, 1131 (2000).
\bibitem{amg} A. M. Green, Phys. Rev. D {\bf 63}, 043005 (2001).
\bibitem{gg} J. E. Gunn and J. R. Gott, Astrophys. J. {\bf 176}, 1 (1972).
\bibitem{drift} M. Lehner et. al., astro-ph/9905074, proceedings of
         `2nd International Conference on Dark Matter in Astro and Particle
          Physics', Heidelberg 1998, 767-771.
\bibitem{ck} C. J. Copi, J. Heo and L. M. Krauss, Phys. Lett. {\bf B461},
             43 (1999); C. J. Copi and L. M. Krauss, astro-ph/0009467.
\bibitem{vergados2} J. D. Vergados, hep-ph/0010151 to appear
         in the proceedings of `NANPino-2000, Non Accelerator New Physics in
         Neutrino Observations', Dubna, Russia.
\bibitem{is} J. R. Ipser and P. Sikivie, Phys. Lett. B {\bf 291}, 288 (1992).
\bibitem{fgb} J. A. Filmore and P. Goldreich, Astrophys. J. {\bf 281} 
                 1 (1984);
              E. Bertschinger, Astrophys. J. Suppl. Ser. {\bf 58}, 39 (1985).
\bibitem{stw} P. Sikivie, I. I. Tkachev and Y. Wang, Phys. Rev. Lett. 
            {\bf 75}, 2911 (1995); Phys. Rev. D {\bf 56}, 1863 (1997).   
\bibitem{ks} W. H. Kinney and P. Sikivie, Phys. Rev. D {\bf 61}, 087305 (2000).
\bibitem{sikiras} P. Sikivie, private communication.
\bibitem{sikvalues} P. Sikivie, Nucl. Phys. Proc. Suppl. {\bf 72}, 110 (1999).
\bibitem{q} K. Fushimi et. al. Phys. Rev. C {\bf 47}, R245 (1993);
            G. J. Davies et. al. Phys. Lett. {\bf B322}, 159 (1994);
            P. F. Smith et. al. Phys. Lett. {\bf B379}, 299 (1996).
\bibitem{gelmgond} G. Gelmini and P. Gondolo, hep-ph/0012315.
\bibitem{idd} C. Calcaneo-Roldan and B. Moore, Phys. Rev. D {\bf 62}, 123005
           (2000); L. Bergstr\"om, J. Edsj\"o and C. Gunnarsson, 
            astro-ph/0012346. 

             


\end{references}
\end{document}